\documentclass{iopart}
\usepackage[T1]{fontenc}
\usepackage{graphicx}
\usepackage{url,graphics,epsfig}
\usepackage{amssymb}

\begin{document}

\jl{2}
%
%
%
\def\etal{{\it et al~}}
\expandafter\ifx\csname url\endcsname\relax
  \def\url#1{{\tt #1}}\fi
\expandafter\ifx\csname urlprefix\endcsname\relax\def\urlprefix{URL }\fi
 \providecommand{\eprint}[2][]{\url{#2}}

%
%
%
%
%
%
\setlength{\arraycolsep}{2.5pt}             
%
%
%

\title[Electron-impact excitation of Ne$^{2+}$]{A large-scale R-matrix calculation 
                                                       for electron-impact excitation of the Ne$^{2+}$, O-like ion}

\author{B M McLaughlin$^{1,2}$\footnote[1]{Corresponding author, E-mail: b.mclaughlin@qub.ac.uk},
             Teck-Ghee Lee$^{3}$, J A Ludlow$^{3}$, E Landi$^{4}$, 
             S D Loch$^{3}$, M S Pindzola$^{3}$ and C P Ballance$^{3}$\footnote[2]{Corresponding author, E-mail: ballance@physics.auburn.edu},}

\address{$^{1}$Centre for Theoretical Atomic, Molecular and Optical Physics (CTAMOP),
                          School of Mathematics and Physics, The David Bates Building, 7 College Park,
                          Queen's University Belfast, Belfast BT7 1NN, UK}

\address{$^{2}$Institute for Theoretical Atomic and Molecular Physics,
                          Harvard Smithsonian Center for Astrophysics, MS-14,
                          Cambridge, MA 02138, USA}

\address{$^{3}$Department of Physics, Auburn University, Auburn, AL 36849, USA}

\address{$^{4}$Department of Atmospheric, Oceanic and Space Sciences, University of
			Michigan, Ann Arbor MI 48109, USA}
%
%

\begin{abstract}

The five J$\Pi$ levels within a $np^2$ or $np^4$ ground state complex
provide an excellent testing ground for the comparison of theoretical line ratios 
with astrophysically observed values, in addition to providing valuable electron
temperature and density diagnostics. The low temperature nature of the 
line ratios ensure that the theoretically derived values are sensitive 
to the underlying atomic structure and electron-impact excitation rates. 
Previous R-matrix calculations for the O-like Ne ion, Ne$^{2+}$,  exhibit spurious 
structure in the cross sections at higher electron energies, which 
may affect Maxwellian averaged rates even at low temperatures. Furthermore,
there is an absence of comprehensive excitation data between the excited states 
that may provide newer diagnostics to compliment the more established 
lines discussed in this paper. To resolve these issues, we present both a
 small-scale 56-level Breit-Pauli (BP)
calculation and a large-scale 554 levels R-matrix Intermediate Coupling Frame Transformation 
(ICFT) calculation that extends the scope and validity of earlier 
JAJOM calculations both in terms of the atomic structure and scattering cross sections. 
Our  results provide a comprehensive electron-impact excitation 
data set for all transitions to higher $n-$ shells. The fundamental 
atomic data for this O-like ion is subsequently used within a collisional radiative framework 
to provide the intensity line ratios across a range of electron temperatures and 
densities of interest in astrophysical observations.        
\end{abstract}



\vspace{1.0cm}
\begin{flushleft}
Short title: Electron impact excitation of Ne$^{2+}$ ions\\
Revised:  J. Phys. B: At. Mol. \& Opt. Phys: \today
\end{flushleft}


\maketitle
%
%
%
%

\section{Introduction}
Electron-impact excitation rates are a fundamental component  
of collisional radiative modelling. In particular, the excitation rates for O-like Ne (Ne$^{2+}$), 
the focus of this paper, are applicable
in diagnosing the spectral lines not only from astrophysical 
observations but also from fusion plasma studies. It has been shown that transitions within the 
first five levels of an atom with the $np^4$ ground configuration can provide 
a valuable diagnostic of the electron temperature \cite{derobertis}
and the electron density \cite{keenan}.
Recently, an updated electron-impact excitation calculation
of Munoz {\it et al.}~\cite{munoz} for the $3s^23p^4$ ground state complex 
of O-like Ar (Ar$^{2+}$) found that the plasma diagnostics
 were particularly sensitive to near threshold resonances, and therefore,
 Ne$^{2+}$ may offer a similar diagnostic potential, but without 
 the high sensitivity to single near threshold resonance as found in the 
 Ar$^{2+}$ case. 
 
 Neon plays an important role in a variety of astrophysical phenomena, being
 the most abundant rare gas in the universe after helium. 
 Forbidden Ne III lines have been observed in a variety of astrophysical plasmas. 
 The lines at 36 $\mu$m and 3869 \AA, due to forbidden 
$\rm ^3P_0 \rightarrow ^3P_1$ and $\rm ^1D_2 \rightarrow~^3P_2$ transitions within 
the ground configuration, have been observed in H II regions by a number of authors 
 \cite{simpson, rubin, baldwin, colgan}.
The Ne III 3869 \AA~ line is also present in Ne-rich filaments in the Supernova Cas A 
remnant \cite{minkowski,fesen} and can be used to investigate the neon 
abundance and its spatial distribution throughout the remnant. 
 
In terms of laboratory measurements, the identification of Ne III lines were 
reviewed by Persson {\it et al.}~\cite{persson} and the low lying terms are
depicted in Fig~1. The present R-matrix calculation now includes shells of 
sufficiently high principal quantum numbers for a direct comparison with 
significantly more of their listed lines. More recently, James and co-workers 
\cite{james} measured electron-impact excitation cross sections for
transitions in the far ultraviolet wavelength range between 1200 and 2700
\AA~using an electron-ion impact collision chamber. 
Daw {\it et al.}~\cite{daw} have measured a pair of lines within 
the ground-state complex. The spin forbidden line $^1$S$_0$$\rightarrow$$^3$P$_1$ 
and the quadrupole $^1$S$_0$$\rightarrow$$^1$D$_2$ line offer diagnostic potential.  
 Finally, Fig~2 illustrates the transitions of interest in the 
planetary nebulae NGC 3918 \cite{clegg} for Ne~III. The wavelength of the 
astrophysically observed transitions are labelled in units of Angstroms. 
     
 An overview of the various theoretical models investigating electron impact excitation of Ne$^{2+}$ is 
summarized in the introduction of McLaughlin and Bell \cite{mclaughlin1}. 
This includes the work of Seaton \cite{seaton1,seaton2,seaton3} in the 1950s, 
and continues through a progression of the perturbative calculations
(e.g., Coulomb-Born \cite{blaha} and distorted wave \cite{czyzak} in 1960s-1970s) 
yet only for transitions within the $1s^22s^22p^4$ ground state complex.
These were followed by early R-matrix calculations of Pradhan \cite{pradhan} 
and Butler and Mendoza \cite{butler} using a program 
of Saraph \cite{saraph1,saraph2} to transform LS-coupled K-matrices 
into level-level cross sections using the {\sc JAJOM} code. This work included six states 
in the close coupling expansion and clearly illustrated the 
sensitivity of excitation rates when adopting either theoretically 
calculated or experimentally observed energy level values. 

Between 1999 and 2001, McLaughlin and co-workers in a couple of 
papers \cite{mclaughlin1,mclaughlin2} extended this model to 28 LS terms or 
56 levels arising from the following $1s^22s^22p^4, 2s2p^5, 2s^22p^33l$($l$=0$-$2), $2s^22p^34s$
configurations, with the extra $\bar{4p}$, $\bar{4d}$ orbitals used only as 
an improvement in the description of the target's states. However, as evident in Fig.~5 
of Ref.\cite{mclaughlin1} there is a breakdown in their calculation 
for the 28 terms model, manifesting as broad unphysical 2~eV resonances 
in transitions from the $^3$P$_0$ ground state term. Unsurprisingly, 
this leads to corresponding unphysical increase in the effective collision 
strength (see Fig.~6 in Ref.\cite{mclaughlin1}) at approximately 
$10^{4.5}$~Kelvin. We point out that a less elaborate six-level model does 
not exhibit the same problem as indicated in Ref.\cite{mclaughlin1}. 

Therefore, it would not be unreasonable, given
the increase in computational power over the last decade to pursue a full
Breit-Pauli R-matrix calculation for the 56 levels under consideration that 
would resolve and correct this problem. However, our present studies revealed 
that the 6 configurations listed above, actually give rise to a possible 49 terms 
or 95 levels. This imbalance between the configuration-interaction of the 
target and those levels actually used in the close coupling expansion will give 
rise to secondary, but a less pronounced set of pseudo-resonances above 3 Rydbergs. 
Therefore, to correct and also expand the scope of Ne$^{2+}$ R-matrix calculations 
currently in the literature we have carried out an Intermediate Coupling 
Frame Transformation (ICFT) \cite{griffin} calculation for 554 levels. 
As illustrated in Ref.\cite{munoz}, the use of the ICFT method gives cross sections
that are extremely close to a Breit-Pauli (BP) calculation if both employing
the same target configurations.   
      
The remainder of the paper is organized as follows. In section~2 we 
outline our small Breit-Pauli R-matrix and a large scale level ICFT R-matrix
calculation used in the current study. Section~3, gives results for selected collision strengths
and effective collision strengths and compares the BP and ICFT calculations.
A small collisional radiative calculation is then performed in section~4, to predict observed 
astronomical line ratios. Finally, in the summary, we assess the 
impact of this now complete data set on the predictive modelling 
of Ne$^{2+}$ line ratios.  

\section{Details of calculations}
In the smaller R-matrix Breit-Pauli model, the atomic orbitals employed  
were generated from the atomic structure code CIV3 \cite{hibbert};
in which the orbitals were energy-optimized on the lowest 28 terms; 
details can be found in the earlier work of McLaughlin and co-workers \cite{mclaughlin1,mclaughlin2}.  
A subset of 28 LS terms constructed from the following configuration list; 
$1s^22s^22p^4$, $1s^22s2p^5$, $1s^22s^22p^33\ell$ and $1s^22p^6$
give rise to 56 fine-structure levels which were used in the close-coupling 
expansion. Extra diffuse 4${\bar p}$ and 4${\bar d}$ pseudo-orbitals were used 
only to improve the target states, but also ensured a larger 
R-matrix box of 15.2 a.u. A total of forty basis orbitals 
were used to represent the continuum basis, easily spanning 
the energy range from threshold to 20 Rydbergs.

In the larger ICFT calculation,
the target radial wavefunctions were generated using GASP (Graphical
Autostructure Package) \cite{gasp}, which is a java front-end for the
atomic-structure code AUTOSTRUCTURE \cite{c11}.
The orbitals were generated within an 
Thomas-Fermi-Dirac-Amaldi potential, using slightly modified lambda scaling
parameters as given by Landi and Bhatia \cite{landi}.
Specifically we used $\lambda_{1s}=1.0$, $\lambda_{2s}=1.3$, $\lambda_{2p}=1.09$, 
$\lambda_{3s}=1.13$, $\lambda_{3p}=1.15$, $\lambda_{3d}=1.11$, 
$\lambda_{4s}=1.12$, $\lambda_{4p}=1.1$, $\lambda_{4d}=1.11$ and $\lambda_{4f}=1.06$.
  However, we are able to include all 24 configurations : $2s^22p^4$, $2s2p^5$, $2p^6$,  $2s^22p^33s$, 
$2s^22p^33p$, $2s^22p^33d$, $2s^22p^34s$, $2s^22p^34p$, $2s^22p^34d$, $2s^22p^34f$,
$2s2p^43s$, $2s2p^43p$, $2s2p^43d$, $2s2p^44s$, $2s2p^44p$, $2s2p^44d$, $2s2p^44f$, 
$2p^53s$, $2p^53p$, $2p^53d$, $2p^54s$,  $2p^54p$, $2p^54d$ and $2p^54f$ 
resulting in 554 levels; which were subsequently
included in close-coupling expansion of the scattering calculation. The energies of the 
low lying $2s2p^5$ terms were much improved over the earlier reported CIV3 values  \cite{hibbert},
 with only the first $^1$D$_2$ level at a 9.6$\%$ difference being the 
greatest outlier relative to NIST energy level values \cite{nist}. 
A representative sample of the full 554 level calculation,
is illustrated in Table 1,  where it is seen that the first 33 level energies agree 
to within an average of $1.26\%$ with the NIST values. 

\begin{table}
\caption{A sample comparison of  our theoretical energies from the 554-level model,  in Rydbergs,
	       for the first 33 levels of Ne III with the NIST \cite{nist} tabulated values.}
\begin{tabular}{ccccccc}
\br
Level No. & Configuration &  Term$^{\rm a}$ & Energy  		& Energy  & $\Delta (\%)^{\rm b}$ & Th. Order \\
		&			&			     &	(Experiment)	&(Theory)	 &				& 		\\	
\mr
1  & 2s$^2$2p$^4$ &  $^3$P$_2$     & 0.00000 & 0.00000 & 0.00             &  1 \\  
2  & 2s$^2$2p$^4$ &  $^3$P$_1$     & 0.00585 & 0.00569 & 2.74		  &  2 \\
3  & 2s$^2$2p$^4$ &  $^3$P$_0$     & 0.00839 & 0.00837 & 0.24		  &  3 \\ 
4  & 2s$^2$2p$^4$ &  $^1$D$_2$     & 0.23548 & 0.25804 & 9.58 		  &  4 \\
5  & 2s$^2$2p$^4$ &  $^1$S$_0$     & 0.50806 & 0.49160 & 3.24		  &  5 \\
6  & 2s$^1$2p$^5$ &  $^3$P$_2$     & 1.86163 & 1.91958 & 3.11		  &  6 \\
7  & 2s$^1$2p$^5$ &  $^3$P$_1$     & 1.86694 & 1.92512 & 3.12		  &  7 \\
8  & 2s$^1$2p$^5$ &  $^3$P$_0$     & 1.86987 & 1.92791 & 3.10		  &  8 \\
9  & 2s$^1$2p$^5$ &  $^1$P$_1$     & 2.63792 & 2.80123 & 6.19		  &  9 \\
10 & 2p$^3$3s     &  $^5$S$_2$     & 2.82384 & 2.81852 & 0.19		  & 10 \\
11 & 2p$^3$3s     &  $^3$S$_1$     & 2.91087 & 2.91399 & 0.11		  & 11 \\
12 & 2p$^3$3p     &  $^5$P$_1$     & 3.17481 & 3.15680 & 0.57		  & 12 \\
13 & 2p$^3$3p     &  $^5$P$_2$     & 3.17509 & 3.15720 & 0.56		  & 13 \\
14 & 2p$^3$3p     &  $^5$P$_3$     & 3.17558 & 3.15781 & 0.56		  & 14 \\
15 & 2p$^3$3s     &  $^3$D$_3$     & 3.21799 & 3.21957 & 0.05		  & 17 \\
16 & 2p$^3$3s     &  $^3$D$_2$     & 3.21825 & 3.21940 & 0.04		  & 16 \\
17 & 2p$^3$3s     &  $^3$D$_1$     & 3.21844 & 3.21934 & 0.03		  & 15 \\

18 & 2p$^3$3p     &  $^3$P$_1$     & 3.25096 & 3.26061 & 0.30		  & 18 \\
19 & 2p$^3$3p     &  $^3$P$_0$     & 3.25105 & 3.26065 & 0.30		  & 19 \\
20 & 2p$^3$3p     &  $^3$P$_2$     & 3.25106 & 3.26075 & 0.30		  & 20 \\

21 & 2p$^3$3s     &  $^1$D$_2$     & 3.26169 & 3.26701 & 0.16		  & 21 \\

22 & 2p$^3$3s     &  $^3$P$_1$     & 3.41259 & 3.37052 & 1.23		  & 23 \\
23 & 2p$^3$3s     &  $^3$P$_0$     & 3.41260 & 3.37048 & 1.23		  & 22 \\
24 & 2p$^3$3s     &  $^3$P$_2$     & 3.41266 & 3.37064 & 1.23		  & 24 \\

25 & 2p$^3$3s     &  $^1$P$_1$     & 3.46123 & 3.42538 & 1.04		  & 25 \\

26 & 2p$^3$3p     &  $^1$P$_1$     & 3.53537 & 3.51826 & 0.48		  & 26 \\

27 & 2p$^3$3p     &  $^3$D$_1$     & 3.54523 & 3.52981 & 0.43		  & 27 \\
28 & 2p$^3$3p     &  $^3$D$_2$     & 3.54533 & 3.53017 & 0.43		  & 28 \\
29 & 2p$^3$3p     &  $^3$D$_3$     & 3.54596 & 3.53137 & 0.41		  & 29 \\

30 & 2p$^3$3p     &  $^3$F$_2$     & 3.56670 & 3.55916 & 0.21		  & 30 \\
31 & 2p$^3$3p     &  $^3$F$_3$     & 3.56684 & 3.55965 & 0.20		  & 31 \\
32 & 2p$^3$3p     &  $^3$F$_4$     & 3.56703 & 3.56029 & 0.19		  & 32 \\

33 & 2p$^3$3p     &  $^1$F$_3$     & 3.57947 & 3.57865 & 0.02		  & 33 \\
\mr
\end{tabular}
 $^a$ $^{2S+1}L_J$\\
$^b$ absolute percentage difference relative to NIST values\\
\end{table}

The advantage of applying the ICFT method is that the R-matrix inner region 
is essentially carried out in LS coupling, including only the mass-velocity 
and Darwin terms. Since we are primarily focusing on line ratios between 
transitions within the ground state complex, we used 
only 20 continuum basis functions for each continuum angular momenta to keep 
the dimensions of the Hamiltonian matrix to a more manageable size. The 
inclusion of the $4p$, $4d$ and $4f$ as spectroscopic orbitals did increase the 
extent of the R-matrix box to 22.88 a.u, an increase of over 50$\%$ in comparison to 
the Breit-Pauli value. The scattering calculations were then performed 
with our set of parallel $R$-matrix programs \cite{dario, connor}, which 
are modified versions of the serial RMATRIX I programs \cite{keith}. 

Partial waves with total angular momentum from $L=0-35$ were calculated, with the Burgess-Tully 
top-up method used to account for higher partial wave contributions. However, 
considering the temperatures of astronomical interest here this should be 
more than sufficient to converge our excitation rates. 

In the outer region, the calculation of the cross sections was split 
into two energy regions. Region 1 spans the energy range from the 
first excitation threshold up to the ionization threshold, whereas region 
2 spans the energy range from from the ionization limit 
to four times the threshold. In region 1, for
both the BP and ICFT models, we 
used a fine energy mesh of 40,000 points 
to resolve the fine Rydberg resonance structure. To account for the limited contribution 
of the higher  partial waves to the low energy cross sections, we interpolated a coarse 
mesh of 200 energy points and added this to the fine mesh cross section. 
In region 2, above the ionisation limit, where a coarse mesh is applicable, only
 200 energy points were also used for both the exchange and
non-exchange calculations. The resulting total collision strengths were subsequently 
Maxwellian averaged across a range of electron temperatures and the 
effective collision strengths \cite{werner} archived for modelling. For comparison with the 56 
level BP calculation, the collision strength file of the 554 level ICFT model was
reduced to include only those transitions between the lowest 56 levels.    

\section{Collision strengths and effective collision strengths}
 
In Fig.~3, a representative set of collision strengths for 
transitions within the ground state $2s^22p^4$ multiplet is given.
It has been shown in previous studies, such as those of Griffin {\it et al.} \cite{griffin} 
and Munoz {\it et al.} \cite{munoz} that both ICFT and BP calculations 
for atomic ions should provide essentially identical results, provided that the number 
of terms/levels included in the configuration-interaction and 
close-coupling expansions are the same.   

However, this is evidently not the case for the BP calculation   
shown in Fig.~3, as although the background collision strengths
between the ICFT and the BP are similar, there are 
spurious resonances in the BP calculation, especially at 
approximately 10 Rydbergs.  In fact, perhaps less obviously, 
for the BP model there should not be any resonance 
structure above 3.2 Rydbergs, as that is the last threshold included 
in that model. Such spurious resonances in the BP calculations would need to be 
removed if reliable rates are to be produced.  
Lastly, it should also be noted that the BP background collision 
strength is higher than ICFT.  

The cumulative effect of these differences is reflected in the
corresponding effective collision strengths \cite{werner} shown in Fig.~4. 
Over a wide range of electron temperatures, the 56 level BP 
effective collision strengths are consistently higher than 
the benchmark 554 level ICFT values. 
However, in terms of line ratios, a low density model using the coronal approximation 
involves the ratio of the effective collision strengths, 
which can lead to smaller than expected differences in the predicted 
line ratios, see the results in section 4.

Figure 5 shows the 2s$^2$2p$^3$3s($^3$S$_{1}$)$\rightarrow$ 
2s$^2$2p$^3$3p($^3$P$_{2,1,0}$) collision strengths as a function of 
electron energy. In this case, although small differences can be seen, the
554 levels ICFT and 56 levels BP calculations are in reasonable accord with each other.
This level of agreement (i.e., $\sim$15$-$25\%) was expected and typical for 
the collision strengths calculated as more target states are included in 
the scattering calculation. As anticipated, in Fig.~6, the corresponding 
effective collision strengths are indeed in good agreement between the 
two calculations.

\section{Emissivity Modelling}
Temperature and density sensitive energy intensity ratios between transitions among the
ground state $3s^23p^4$ multiplet have been studied for Ar~III \cite{keenan,munoz}. 
Here we examine the corresponding energy intensity ratios for transitions among the ground 
state $2s^22p^4$ multiplet of Ne~III. The effective collision strengths are processed 
through the ADAS suite of codes (\url{http://www.adas.ac.uk}) with the first 5 levels 
in both the 56 level BP and 554 level ICFT calculation,  
moved to NIST energies and NIST Einstein A-coefficients where available.
The NIST values are taken primarily from the work of Froese Fischer and Tachiev  
\cite{charlotte}  and Kramida and Nave \cite{kramida}. The wavelengths for the various 
transitions used in the emissivity modelling are given in figure 4.

For the temperatures and densities of interest, cascades from higher levels are not 
significant. Therefore, although both the ICFT and BP collisional rates include 
those involving the first 56 levels, only the transitions within the ground state 
complex are important for the following line ratios. Both datasets used the same 
A-values for transitions within the ground configuration. 

We first examine the temperature sensitive energy intensity ratio
\begin{equation}
R_1=\frac{(N_4 A_{4\rightarrow 1}/\lambda_{4\rightarrow 1})+(N_4 A_{4\rightarrow 2}/\lambda_{4\rightarrow 2})}
{N_5 A_{5\rightarrow 4}/\lambda_{5\rightarrow 4}}
\end{equation}
with $N_i$ the population in energy level $i$, $A_{i\rightarrow j}$ the Einstein A coefficient for a transition
between energy levels $i$ and $j$ and $\lambda_{i\rightarrow j}$ the corresponding wavelength for the transition. 
Note that the index numbers for the levels are given in Table 1.

The BP and ICFT results are shown in Fig. 7. The BP and ICFT results are in good agreement with each other even though the
BP effective collision strengths for these transitions were higher than the ICFT results. This is due 
to a cancellation of errors in the ratio. Due to the fact that the BP effective collision strengths for
transitions within the ground configuration are in general higher than
the ICFT ones, the BP model has increased emissivities for all of the
transitions.  While this does not result in a large change in the line
ratios, one would see a difference when looking at absolute intensities
of individual spectral lines. The results from the BP model provides larger
emissivities which can affect measurements of the Ne elemental abundances,
since such measurements are obtained using the absolute emissivity, or
the ratio with the emissivity of a completely different ion/element. The
measurements based on the BP results would lead to a lower Ne abundance. 

Next we consider the density sensitive energy intensity ratio
\begin{equation}
R_2=\frac{N_4A_{4\rightarrow 1}/\lambda_{4\rightarrow 1}}
{N_2A_{2\rightarrow 1}/\lambda_{2\rightarrow 1}}
\end{equation}
with the BP and ICFT results shown in Fig. 8. The $R_2$ ratio for the BP and ICFT results agree closely with each 
other and show a strong density dependence between electron densities of $10^4$ and $10^8$ cm$^{-3}$.
As for the $R_1$ ratio, the $R_2$ ratio for Ne~III shows a wider range of variation than the
$R_2$ ratio for Ar~III. Therefore, although systematically the two systems are both excellent 
density diagnostics, the Ne~III is the marginally better option.  

\section{Summary}

A large scale electron-impact excitation calculation of this O-like ion was carried out 
to investigate high energy irregularities in the cross sections of an
earlier R-matrix calculation and the subsequent impact on line ratios. We have resolved these issues,
and have benchmarked effective collision strengths for use in the modelling of 
laboratory and astrophysical plasmas. As an illustration of an application
of these effective collision strengths, a study was undertaken into the viability of applying  
the same Ar~III line ratios \cite{munoz} to the Ne~III case. Overall, the O-like ion, Ne III, exhibits 
great promise in diagnosing the temperature and density of Ne~III in planetary nebulae plasmas, 
being more sensitive than the equivalent Ar III line ratios.

Our effective collisions strengths for the 554 level ICFT R-matrix calculation shall be made available
on the Oak Ridge National Laboratory Atomic Data website
 (\url{http://www-cfadc.phy.ornl.gov/data_and_codes})
and in the ADAS database (\url{http://www.adas.ac.uk}).

\section{Acknowledgements}
It is a pleasure to thank Professor A Dalgarno FRS, for drawing our attention 
to this problem and the need for accurate atomic data required for the modelling 
of planetary nebulae. T-G Lee, C P Ballance and J A Ludlow were supported 
by US Department of Energy (DoE) grants through Auburn University.
E Landi acknowledges support from NASA grants NNX10AM17G and NNX11AC20G.  
B M McLaughlin acknowledges support by the US
National Science Foundation through a grant to ITAMP
at the Harvard-Smithsonian Center for Astrophysics.
The computational work was carried out at the National Energy Research Scientific
Computing Center in Oakland, CA, USA and on the Tera-grid at
the National Institute for Computational Sciences (NICS) in Knoxville, TN, USA,
which is supported in part by the US National Science Foundation.

%
%
%
%
%

%
%
%
%

\newpage
%
%
%
%

\begin{figure}
\begin{center}
\includegraphics[scale=0.6]{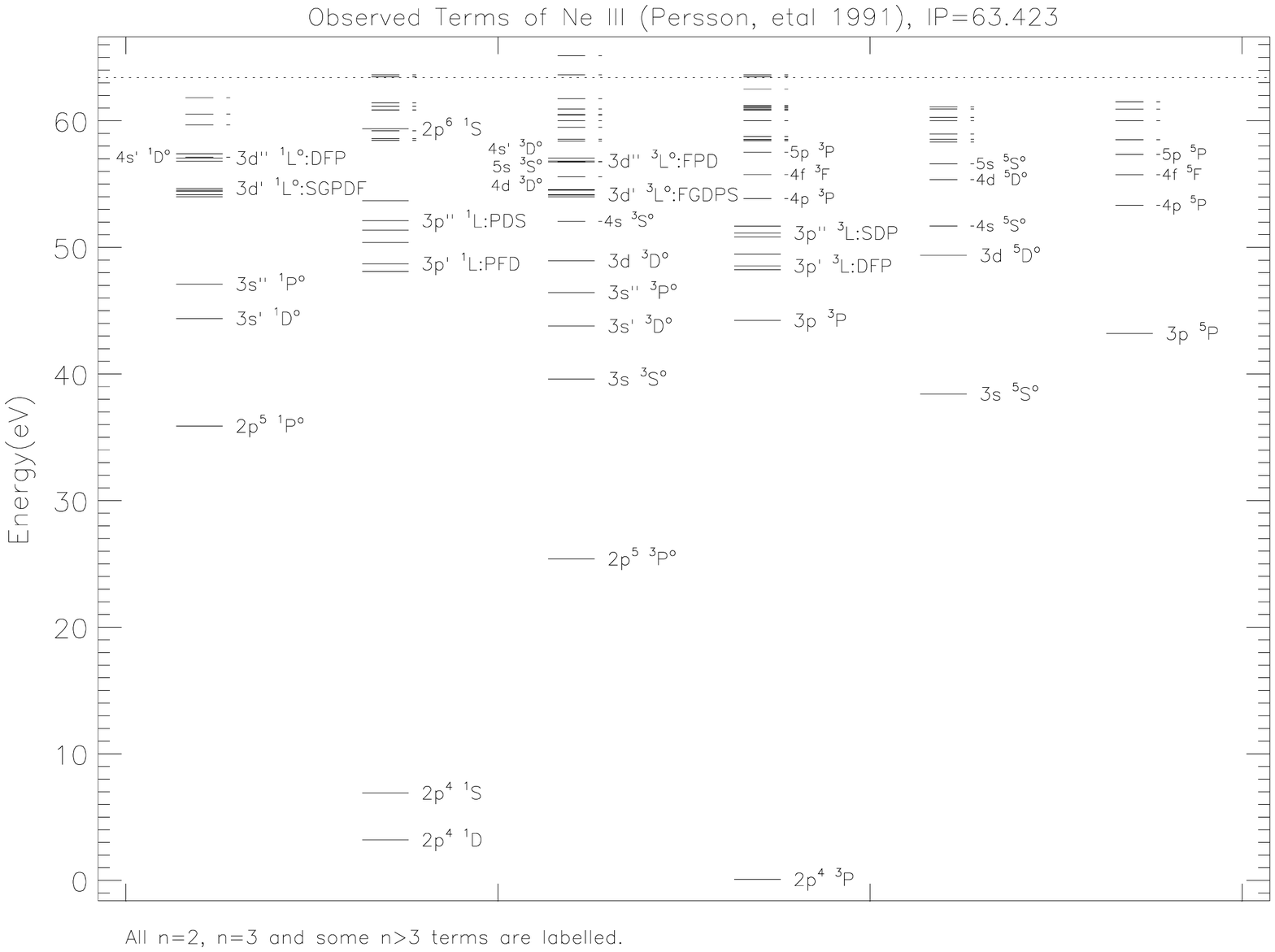}
\caption{\label{fig:Fig1} 
Ne III, energy level diagram (levels given in eV) 
showing the observed low lying levels 
of NeIII from the laboratory work of Persson {\it et al.} 
\cite{persson}. }                                           
\end{center}
\end{figure}

\begin{figure}
\begin{center}
\includegraphics[scale=0.7]{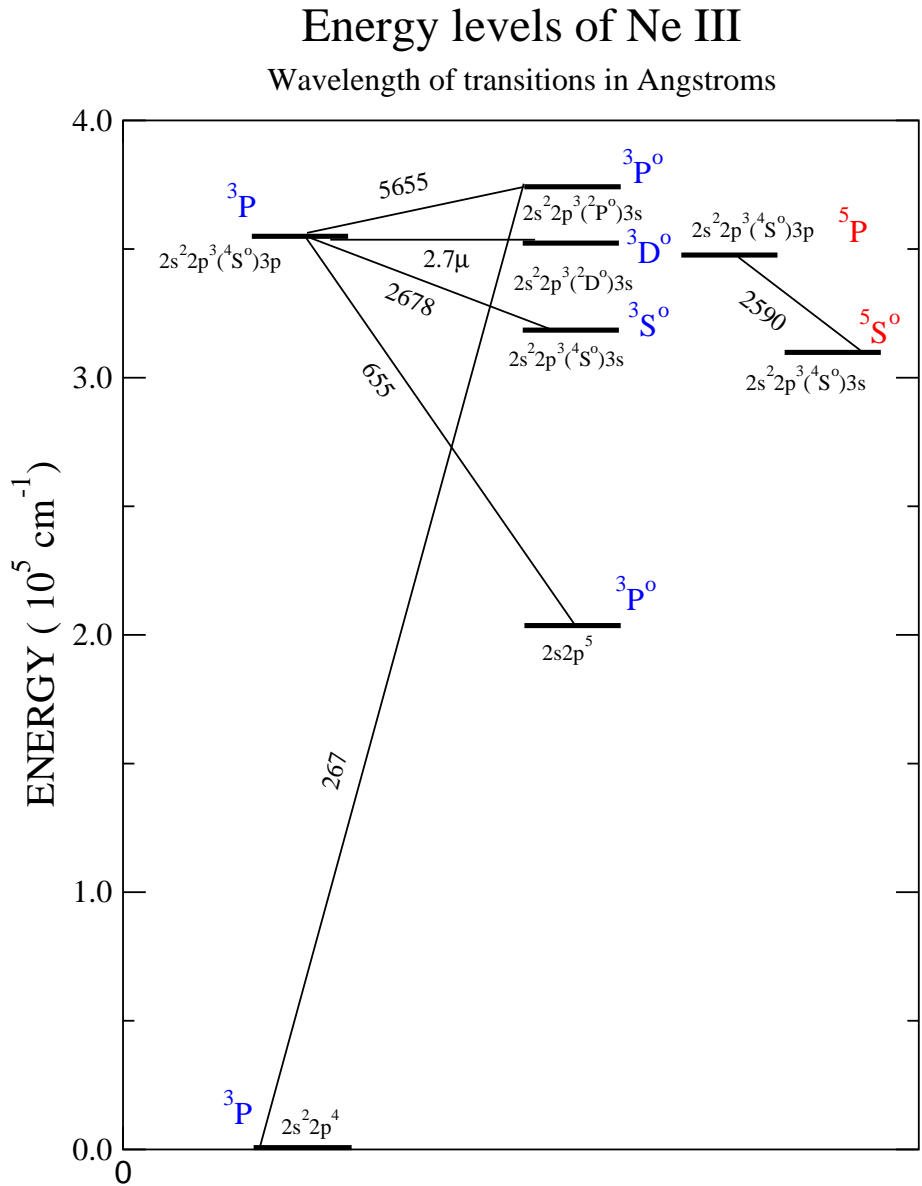}
\caption{\label{fig:Fig2} (Colour online)
An overview of energy levels (levels given 
in cm$^{-1}$) illustrating the transitions of interest in the 
planetary nebulae NGC 3918 \cite{clegg} for Ne~III. The wavelength of the 
astrophysically observed transitions are labelled in units of Angstroms. 
For multiplets, the wavelength (in \AA) of the strongest observed lines is given. 
Singlet levels are omitted for clarity.}
\end{center}
\end{figure}

\begin{figure}
\begin{center}
\includegraphics[scale=0.6]{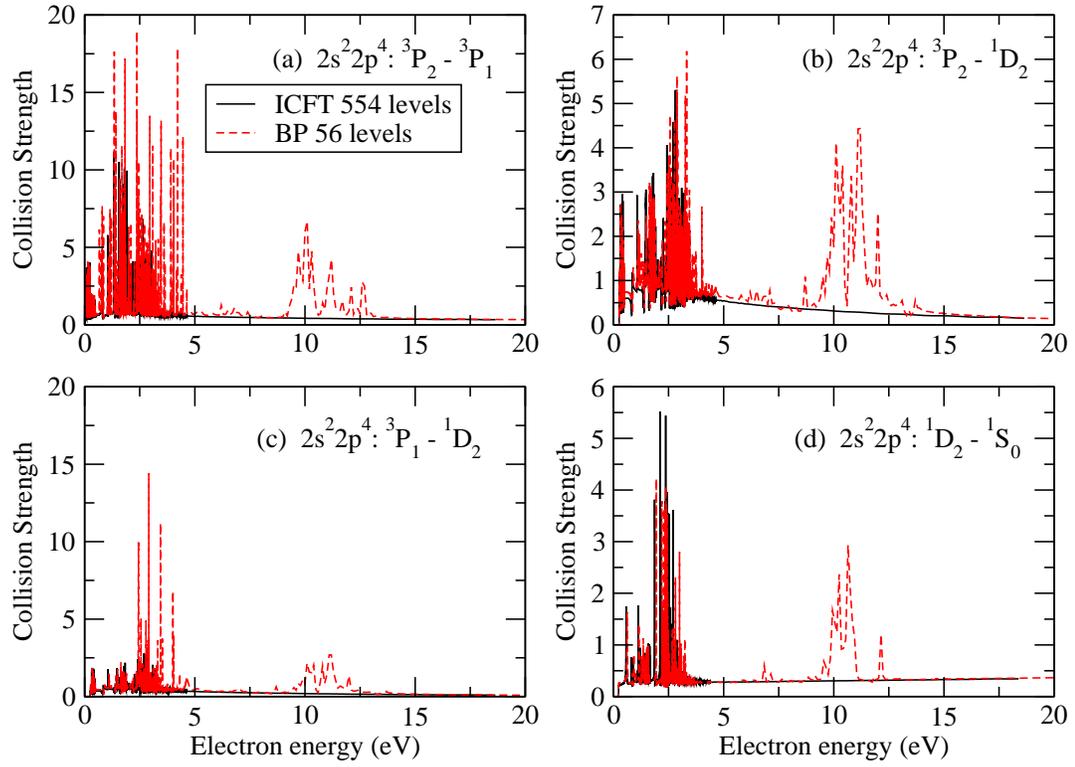}
\caption{\label{fig:Fig3} (Colour online)
Collision strengths for the transitions (a) $\rm 2s^22p^4 : ^3P_2 -  ^3P_1$,
(b) $\rm 2s^22p^4 : ^3P_2 - ^1D_2$, (c) $\rm 2s^22p^3: ^3P_1 -  ^1D_2$ 
and (d) $\rm 2s^22p^4: ^1D_2 -  ^1S_0$
as a function of the incident electron energy. The solid black line 
indicates the 554 level ICFT results and the dashed red  line indicates the 
56 level Breit-Pauli results.}
\end{center}
\end{figure}

\begin{figure}
\begin{center}
\includegraphics[scale=0.6]{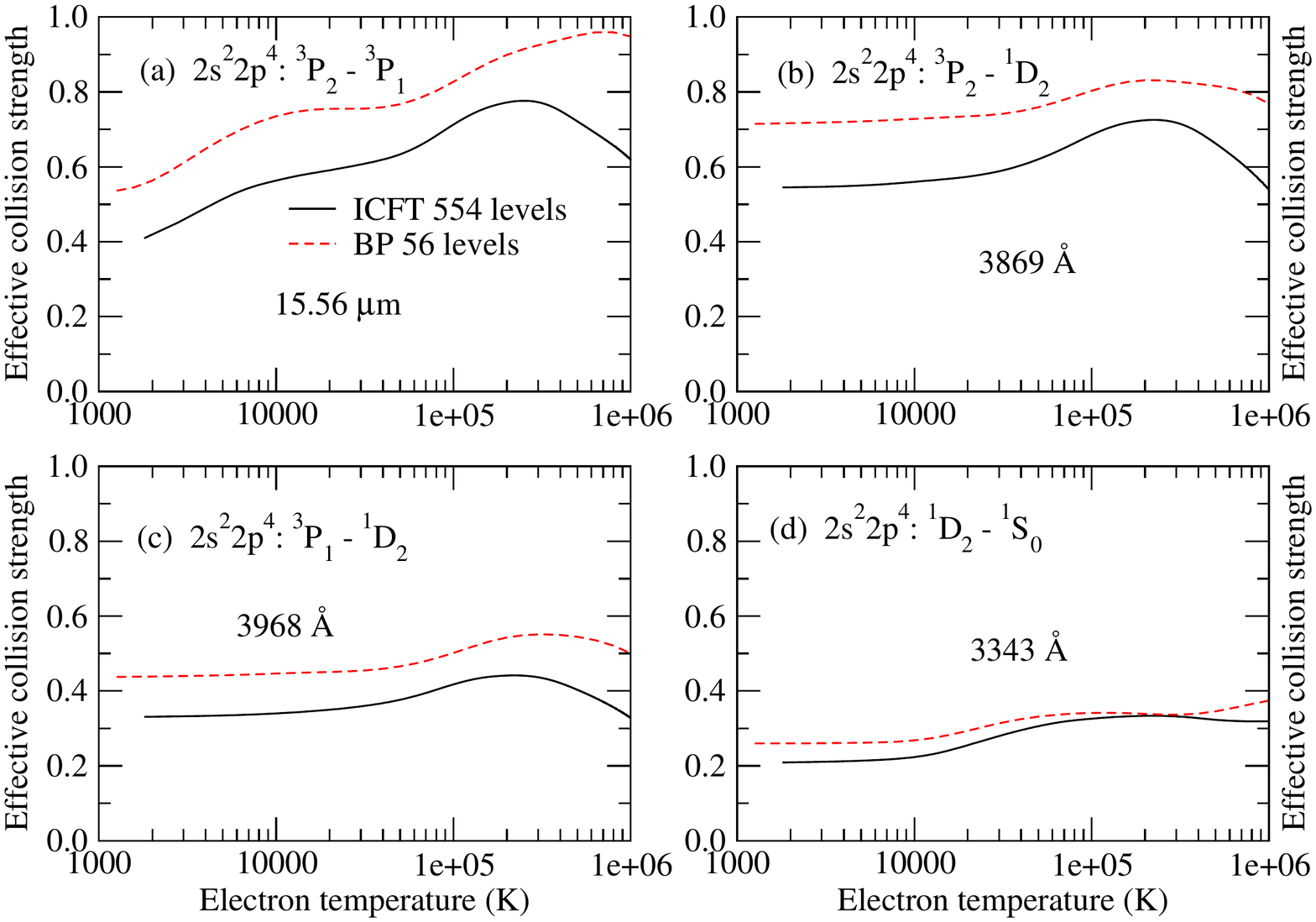}
\caption{\label{fig:Fig4} (Colour online)
Effective collision strengths obtained by averaging the collision strengths over a 
Maxwellian temperature distribution for the electrons.  
Results are shown for the transitions (a) $\rm 2s^22p^4 : ^3P_2 -  ^3P_1$,
(b) $\rm 2s^22p^4 : ^3P_2 - ^1D_2$, (c) $\rm 2s^22p^3: ^3P_1 -  ^1D_2$ 
and (d) $\rm 2s^22p^4: ^1D_2 -  ^1S_0$
as a function of the electron temperature (K). The solid black line 
indicates the 554 level ICFT results and the dashed red line indicates the 
56 level Breit-Pauli results.  The wavelengths in Angstroms for the 
various transitions are also included for completeness}
\end{center}
\end{figure}

\begin{figure}
\begin{center}
\includegraphics[scale=0.6]{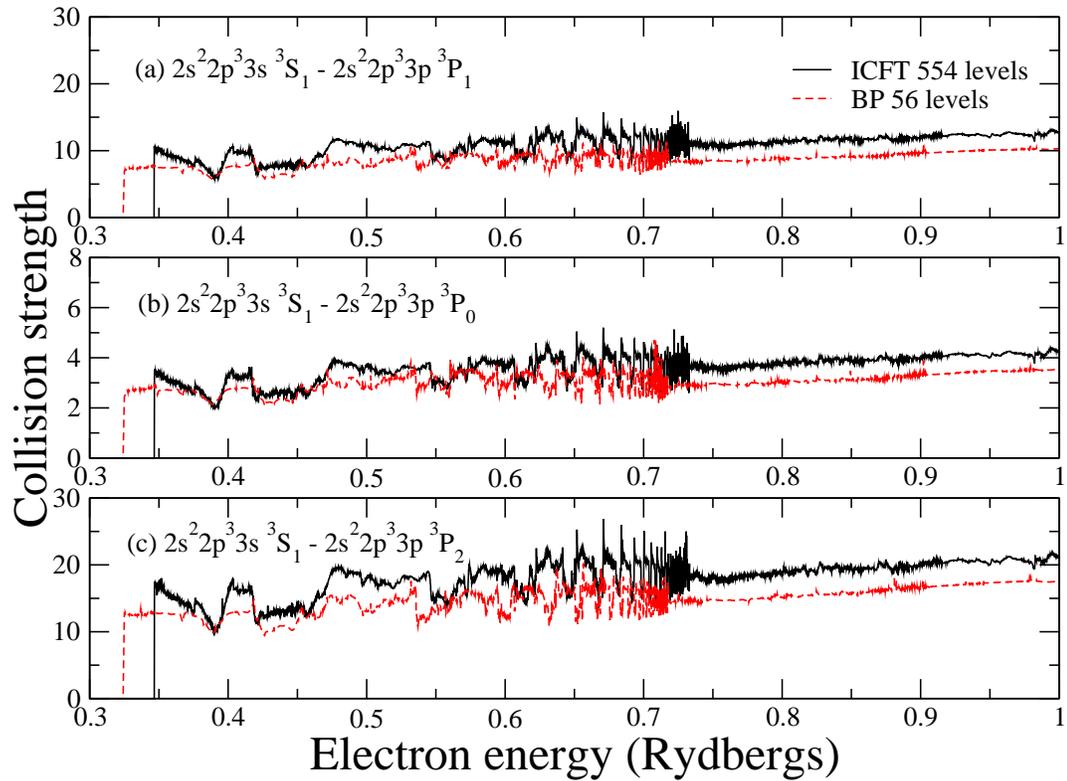}
\caption{\label{fig:Fig5} (Colour online)
Collision strengths for the transitions (a) $\rm 2s^22p^33s ^3S_1 - 2s^22p^33p ^3P_1$,
(b) $\rm 2s^22p^33s ^3S_1 - 2s^22p^33p ^3P_0$, (c) $\rm 2s^22p^33s ^3S_1 - 2s^22p^33p ^3P_2$
as a function of the incident electron energy. The solid black line 
indicates the 554 level ICFT results and the dashed red line indicates the 
56 level Breit-Pauli results.}
\end{center}
\end{figure}

\begin{figure}
\begin{center}
\includegraphics[scale=0.6]{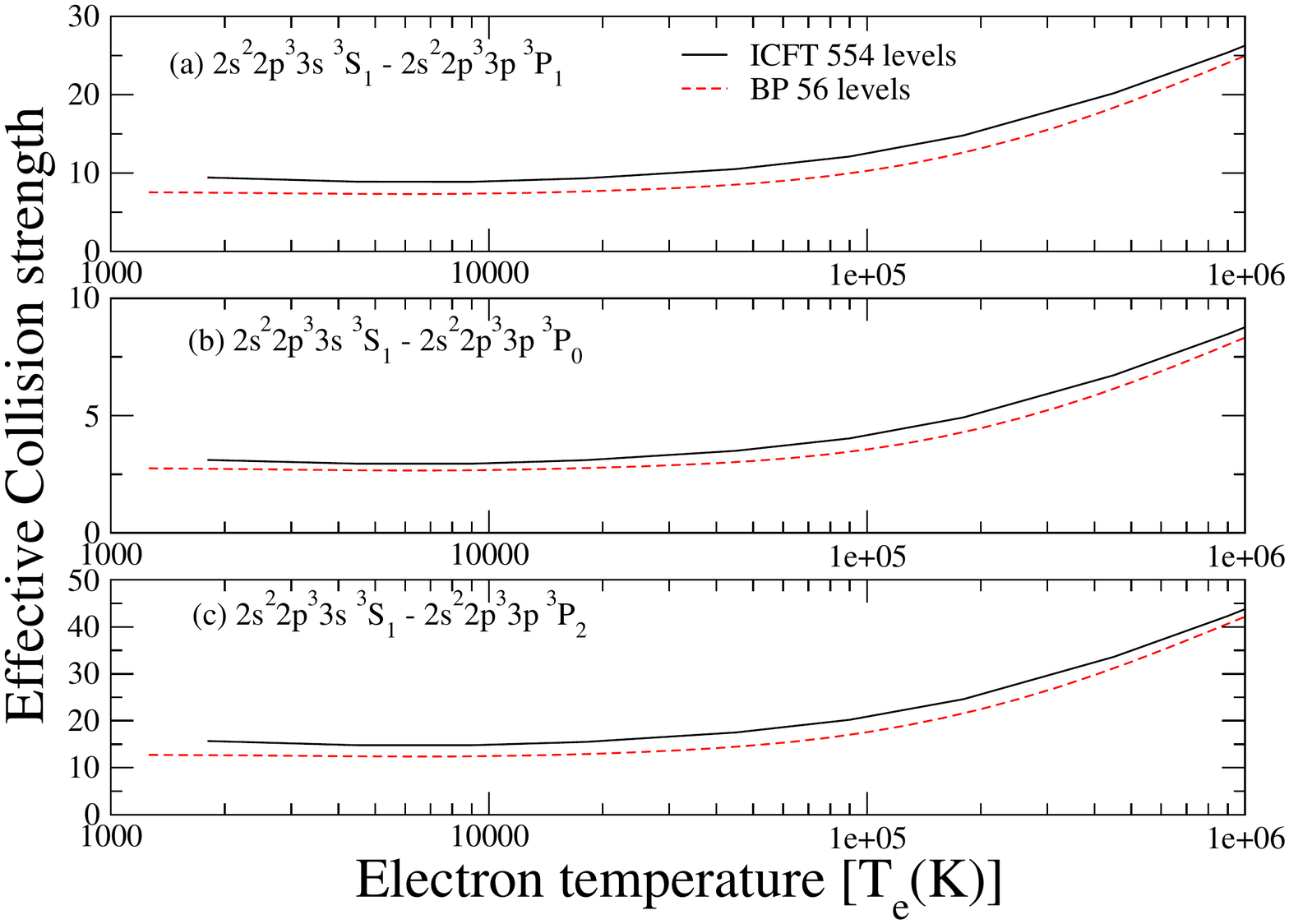}
\caption{\label{fig:Fig6} (Colour online)
Effective collision strengths obtained by averaging the collision strengths over a 
Maxwellian temperature distribution for the electrons.  Results are shown
for the transitions (a) $\rm 2s^22p^33s ^3S_1 - 2s^22p^33p ^3P_1$,
(b) $\rm 2s^22p^33s ^3S_1 - 2s^22p^33p ^3P_0$, (c) $\rm 2s^22p^33s ^3S_1 - 2s^22p^33p ^3P_2$
as a function of the electron temperature (K). The solid black line 
indicates the 554 level ICFT results and the dashed red line indicates the 
56 level Breit-Pauli results.}
\end{center}
\end{figure}

\begin{figure}
\begin{center}
\includegraphics[scale=0.6]{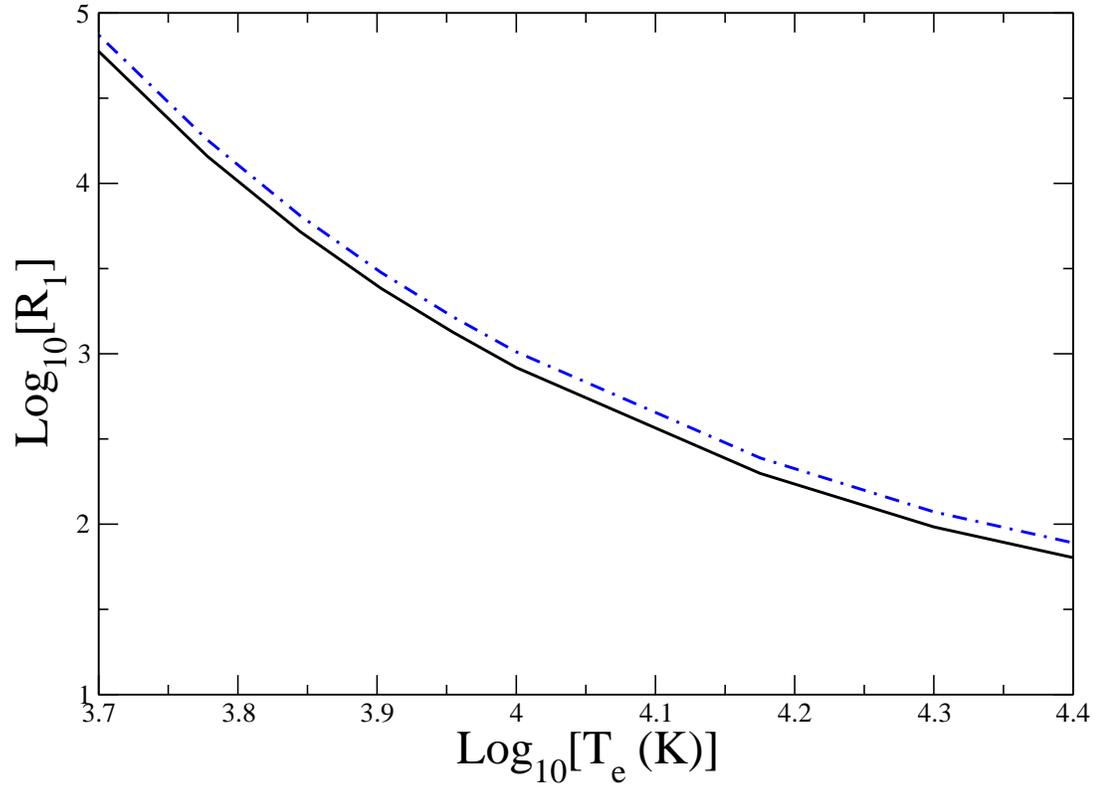}
\caption{\label{fig:Fig7} (Colour online)
$R_1$ line ratio as a function of electron temperature for an electron density
of $N_e=1\times10^3$cm$^3$. The solid black  line shows the results using the 554 level
ICFT R-matrix calculation and the dot-dashed blue line shows the result using the 56 level
Breit-Pauli R-matrix calculation, with the energies and transition rates for the first
5 levels in both R-matrix calculations shifted to NIST energies and transition 
rates\protect\cite{nist} for the modelling calculations.}
\end{center}
\end{figure}

\begin{figure}
\begin{center}
\includegraphics[scale=0.6]{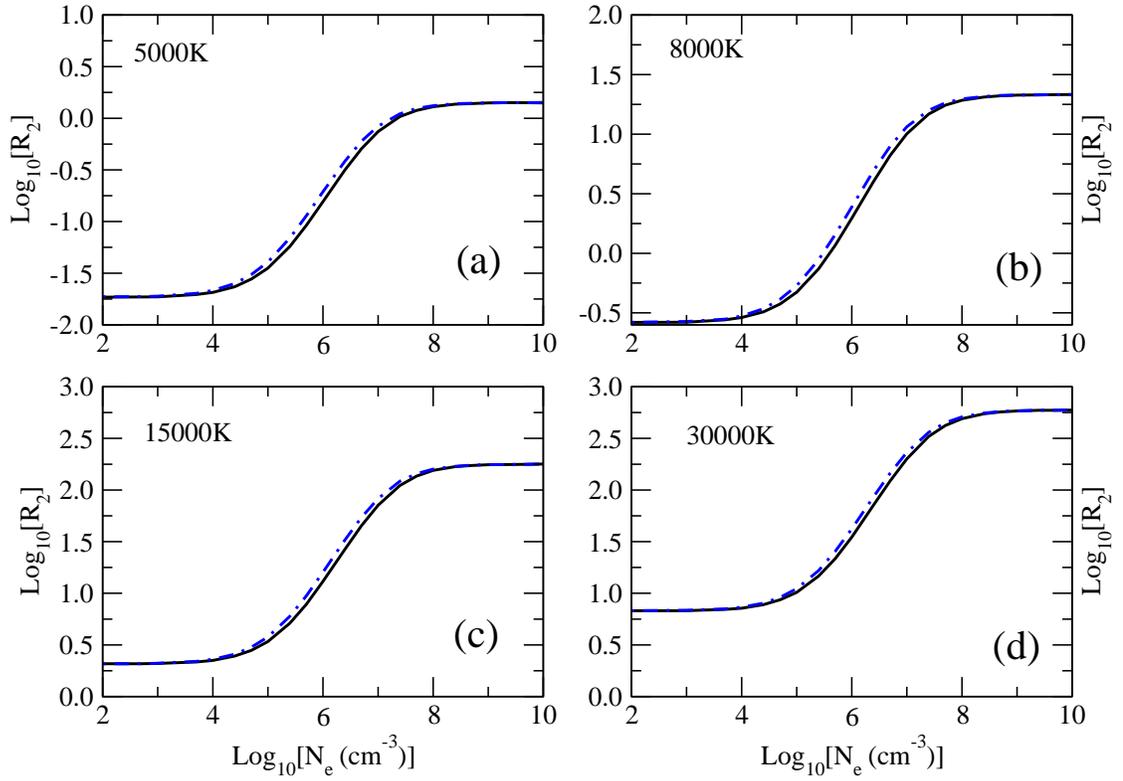}
\caption{\label{fig:Fig8} (Colour online)
$R_2$ line ratio as a function of electron density for electron
temperatures of (a) 5000K (b) 8000K (c) 15000K,  and (d) 30000K. 
The solid black line shows the results using the 554 level
ICFT R-matrix calculation and the dot-dashed blue line shows the result using the 56 level
Breit-Pauli R-matrix calculation.}
\end{center}
\end{figure}

\end{document}